**Quantum Rabi dynamics of trapped atoms far in the deep strong coupling regime**

Johannes Koch,[1] Geram R. Hunanyan,[1] Till Ockenfels,[1] Enrique Rico,[2,3,4] Enrique Solano[3,4,5,6] and Martin Weitz[1]

[1]*Institut für Angewandte Physik, Universität Bonn, Wegelerstr. 8, 53115 Bonn, Germany*

[2]*EHU Quantum Center, University of the Basque Country UPV/EHU, P.O. Box 644, 48080 Bilbao, Spain*

[3]*Department of Physical Chemistry, University of the Basque Country UPV/EHU, Apartado 644, 48080 Bilbao, Spain*

[4]*IKERBASQUE, Basque Foundation for Science, Plaza Euskadi 5, 48009 Bilbao, Spain*

[5]*Kipu Quantum, Greifswalder Straße 226, 10405 Berlin, Germany*

[6]*International Center of Quantum Artificial Intelligence for Science and Technology (QuArtist) and Department of Physics, Shanghai University, 200444 Shanghai, China*

*Correspondence and requests for materials should be addressed to J.K. (email: koch@iap.uni-bonn.de) or to M.W. (email: martin.weitz@uni-bonn.de)*

**The coupling of a two-level system with an electromagnetic field, whose fully quantized version is the quantum Rabi model, is among the central topics of quantum physics. When the coupling strength becomes large enough that the field mode frequency is reached, the deep strong coupling regime is approached, and excitations can be created from the vacuum. Here we demonstrate a periodic variant of the quantum Rabi model in which the two-level system is encoded in the Bloch band structure of cold rubidium atoms in optical potentials. With this method we achieve a Rabi coupling strength of 6.5 times the field mode frequency, which is far in the deep strong coupling regime, and observe a subcycle timescale raise in bosonic field mode excitations. In a measurement recorded in the basis of the coupling term of the quantum Rabi Hamiltonian, a freezing of dynamics is revealed for small frequency splittings of the two-level system, as expected**

**when the coupling term dominates over all other energy scales, and a revival for larger splittings. Our work demonstrates a route to realize quantum-engineering applications in yet unexplored parameter regimes.**

**Introduction**

The motivation to develop the quantum Rabi model [1,2], which is also termed the single-mode spin-boson model, mostly stems from the quest to obtain a complete quantum description of the interaction of matter and light [3,4,5,6], and currently this topic is also highly relevant in the context of quantum information technologies [7,8,9,10,11,12,13]. For increased coupling between matter and light, as the coupling strength becomes stronger than the decoherence rate, the so-called strong coupling regime is reached, with mixed states of the two-level system and the field mode becoming relevant, as can be described in terms of the Jaynes-Cummings model developed earlier [14]. The quantum Rabi model, in addition to the co-rotating, also includes the counter-rotating terms of the interaction Hamiltonian, which has striking consequences as the coupling strength approaches the eigenfrequency of the oscillator, a regime that is not accessible to natural light-matter interactions. On the theoretical side, an analytic solution of the full quantum Rabi model has more recently been found [15]. Experimentally, implementations of the quantum Rabi model using Josephson qubit, metamaterial, and spin-motion cold atom settings have reached values of the ratio of coupling $g$ and bosonic mode frequency $\omega$ of up to 1.43 [16,17,18,19]. A recent ion trapping experiment reporting a quantum phase transition in the ground state dynamics of the quantum Rabi model has operated in a regime with $g/\omega \approx 3.55$ [20].

Using an experimental approach based on implementing the quantum Rabi model in the Brillouin zone of trapped cold atoms, we demonstrate a coupling ratio of $g/\omega \approx 6.5$. A regime where the coupling term dominates over all other energy scales can be experimentally accessed in a wide parameter range. Our approach uses a two-level system provided by two Bloch bands in an optical lattice and a bosonic mode provided by the quantized atomic vibration in a superimposed optical dipole trap potential. For short interaction times, predictions of the quantum Rabi model in the investigated parameter regime are experimentally validated. For long interaction times, upon that the edge of the Brillouin zone is reached we observe the onset of the dynamics of a proposed generalized periodic version of this model [21].

**Results**

**Background and experimental realization.** A schematic of the relevant vibrational modes of ultracold atoms in the implemented potential landscape is shown in Fig. 1a, see the left hand side for an illustration of the bosonic mode represented by the quantized atomic vibration in the harmonic trapping potential. The right hand side shows the superimposed periodic lattice potential serving to implement the two-state system in its Bloch band structure, and atomic wavepackets now evolve in time in the combined potential, see the bottom schematic. Before discussing the degree of coupling between these two quantized modes for typical experimental parameters, we briefly describe our experimental implementation, see Fig. 1b for a schematic of the used setup. A harmonic trapping potential for a cold cloud of rubidium atoms ($^{87}$Rb) is generated by a focused laser beam derived from a $CO_2$-laser operating near a wavelength of 10.6 μm, which due to its large detuning from the atomic resonances allows for the creation of deep potentials while keeping the scattering rate low enough to prevent spurious heating of the atomic ensemble. The additional lattice potential, of spatial periodicity $\frac{\lambda}{4}$, where $\lambda = 783.5$ nm denotes the wavelength of the driving laser beams, is generated by the dispersion of Doppler-sensitive Raman transitions [22,23], see Fig. 1c for the used level scheme. These effective four-photon processes couple atoms in momentum states $|-2\hbar k + q\rangle$ and $|2\hbar k + q\rangle$, where q denotes the atomic quasimomentum and $k = \frac{2\pi}{\lambda}$. The coupling leads to a splitting between bands, see Fig. 1d for the resulting atomic dispersion, and we in the following restrict the discussion to the lowest two bands. At the band crossing (at $q = 0$) we are left with the eigenstates of the two-level system of the quantum Rabi Hamiltonian, with $|g\rangle = \frac{1}{\sqrt{2}}(|-2\hbar k\rangle + |2\hbar k\rangle)$ and $|e\rangle = \frac{1}{\sqrt{2}}(|-2\hbar k\rangle - |2\hbar k\rangle)$ respectively, whose coupling to the bosonic mode provided by the vibrational dynamics we are interested in. While in the general case the system is described by a periodic variant of the quantum Rabi model (see Methods and Ref. 21), which interestingly also maps on a Hamiltonian realizable in a fluxionium superconducting qubit setting [24], we in the present work concentrate on interaction times short enough to remain in the first Brillouin zone such that both the quantum Rabi and the periodic quantum Rabi models coincide. The atomic dynamics in this regime is determined by the quantum Rabi Hamiltonian

$$\widehat{H} = \hbar\omega\, \hat{a}^\dagger \hat{a} + \frac{\hbar\omega_q}{2}\sigma_z + i\hbar g \sigma_x(\hat{a}^\dagger - \hat{a}), \tag{1}$$

where $\hat{a}^\dagger$ and $\hat{a}$ correspond to creation and annihilation operators of the bosonic field, with as usual $\hat{x} = \sqrt{\frac{\hbar}{2m\omega}}(\hat{a} + \hat{a}^\dagger)$, $\hat{q} = \sqrt{-\frac{\hbar m\omega}{2}}(\hat{a}^\dagger - \hat{a})$, and $\sigma_x$ and $\sigma_z$ are Pauli matrices that act on two-component spinors with the components describing course-grain atomic wavefunctions in upper and lower bands respectively, $\sigma_x = |n_b = 0\rangle\langle n_b = 0| - |n_b = 1\rangle\langle n_b = 1|$ and $\sigma_z = |n_b = 1\rangle\langle n_b = 0| + |n_b = 0\rangle\langle n_b = 1|$, with the Bloch band index $n_b$ (see Methods). Further, $\hbar\omega_q$ is the energetic spacing between the bands at the position of the crossing (Fig. 1d), which can be adjusted by the depth of the lattice potential, and $g = k\sqrt{\frac{2\hbar\omega}{m}}$ is the coupling constant. This magnitude of the coupling is well-understood in terms of the energy transfer between momentum picture states $|-2\hbar k + q\rangle$ and $|2\hbar k + q\rangle$ being of order $\Delta E = (q + 2\hbar k)^2/2m - (q - 2\hbar k)^2/2m = 2\hbar k \cdot q/m \approx \sqrt{n}\hbar g$ with the above value of the coupling constant for a typical value of $q \approx \sqrt{n} \cdot \sqrt{m\hbar\omega/2}$ in the harmonic oscillator potential. Typical experimental parameters are a trap oscillation frequency $\frac{\omega}{2\pi} \in [350, 750]$ Hz, for which we arrive at $\frac{g}{2\pi}$ between 2290 Hz and 3090 Hz, so that the deep strong coupling limit is well fulfilled, meaning that the two motional modes present in the system, see also Fig. 1a, exchange energy with each other faster than the temporal period. The qubit frequency spacing $\frac{\omega_q}{2\pi}$ can be tuned between 0 Hz and 5.5 kHz. In the coupling regime of $g \gg \omega$ that we study, striking dynamics is observed at this point.

**Temporal evolution of system excitations.** To begin with, we have characterized the temporal evolution of the bosonic excitation number, as to verify the presence of quantum Rabi physics in the deep strong coupling regime. For this, both the atomic dynamics in position space was monitored by spatial imaging of the atomic cloud following its manipulation in the combined lattice and harmonic dipole trapping potential, as well as the atomic dynamics in momentum space by time-of-flight imaging. In this way, the expectation value for the bosonic excitation number $\langle N \rangle$, where $\hbar\omega \left(\langle N \rangle + \frac{1}{2}\right) = \frac{m\omega^2}{2} \langle x^2 \rangle + \frac{1}{2m} \langle q^2 \rangle$, can be determined. For these measurements, atoms are initially prepared at a momentum centered at $|-2\hbar k\rangle$, corresponding to a quasimomentum of $q = 0$ in the Brillouin zone, see Fig. 1d, and $\langle N \rangle = 0$. The blue dots in Fig. 2a show the observed temporal dynamics of the atomic excitation number in our system for a qubit splitting $\frac{\omega_q}{2\pi} = 586(6)$ Hz, which is of the order of the harmonic trapping frequency $\frac{\omega}{2\pi} = 346(7)$ Hz. The observed increase of the excitation number $\langle N \rangle$ with time shows that the deep strong coupling regime is reached. In general, the experimental data is in good agreement with theory based on numerically integrating the Schrödinger equation using the Hamiltonian of eq. 1, for the large coupling strength of $\frac{g}{\omega} \cong 6.5$ used in the experiment (blue line). Remaining differences visible especially for shorter interaction time are attributed to the limited spatial resolution of the imaging system of 6.5 μm (Methods), causing systematic uncertainties in the determination of the moment $\langle x^2 \rangle$. For comparison, the orange data points correspond to data for the larger qubit spacing of $\frac{\omega_q}{2\pi} = 5200(50)$ Hz, at which for the used value of $\frac{g}{2\pi} = 2275(23)$ Hz the dispersive deep strong coupling regime, defined as $\omega_q \geq g$ [21], is reached. Here, the increase in excitation number occurs more slowly. Next, we have recorded experimental data for different ratios of the relative coupling strength $\frac{g}{\omega}$. For this, the trapping frequency $\omega$ was tuned, and Fig. 2b gives corresponding data recorded at the fixed interaction time of $t = \frac{3}{8}\frac{\pi}{\omega}$ versus the relative coupling strength $\frac{g}{\omega}$ both for a qubit frequency of $\frac{\omega_q}{2\pi} = 590(6)$ Hz (blue dots) and $\frac{\omega_q}{2\pi} = 5850(60)$ Hz (orange triangles) respectively. The data shows that the excitation number increases with the relative coupling strength $\frac{g}{\omega}$, and the achieved large values of up to above 70 excitation quanta, which are achieved at the used short subcycle interaction time, i.e. being much shorter than the period $\frac{2\pi}{\omega}$, gives evidence that we operate in the regime of the coupling strength g far exceeding the oscillator frequency ω. This can be seen analytically when for sake of simplicity as a lower bound the formula for the maximum value of the excitation number in the slow qubit approximation $\omega_q \cong 0$ for which a displaced harmonic oscillator model applies, of $\frac{g}{\omega} \geq \frac{\sqrt{\langle N \rangle}}{2}$ is used (see Methods). For a quantitative comparison, given both that we operate at a nonvanishing value of $\omega_q$ and that at the used interaction times the maximum of the bosonic excitation number is not yet reached, we have to rely on a comparison to a numeric solution of the Hamiltonian; see the good agreement of the experimental data with corresponding theory in the two different regimes.

**Dynamics of real and momentum space mean values.** Next, we have analyzed the variation of the mean displacement $\langle x \rangle$ of the atomic cloud from the trap center versus time. For this measurement, atoms are prepared at momentum of $|-2\hbar k\rangle$ and after evolution in the combined lattice and harmonic trapping potential imaged in real space. Corresponding experimental data is shown in Fig. 3a as a function of interaction time for different values of the qubit frequency $\omega_q$, as tuned by adjusting the lattice depth. For a vanishing depth of the lattice potential, i.e. in the slow qubit limit of $\omega_q \to 0$, we observe the onset of a harmonic oscillation in the harmonic trapping potential, while for increasing lattice depth, corresponding to a non-vanishing value of the qubit spacing $\omega_q$, the observed displacement is reduced, and the evolution for stronger lattice potentials becomes nonharmonic. From the observed displacement of the $\omega_q = 0$ data, we can readily determine the ratio of the coupling to the oscillation frequency, which equals $\frac{g}{\omega} = \frac{x_{m,0}}{x_{\text{ho}}}$, where $x_{m,0} = \frac{2\hbar k}{m\omega}$ is the amplitude of the classical oscillation in the absence of a lattice potential and $x_{\text{ho}} = \sqrt{\frac{2\hbar}{m\omega}} \approx 0.82$ µm the size of the harmonic oscillator ground state wavepacket. From this, we obtain $\frac{g}{\omega} = 5.6(6)$, which is near the above-described result for the coupling ratio. To put the deep strong coupling condition $\frac{g}{\omega} \gg 1$ differently, only in this limit the splitting of wavepackets in the bosonic mode expected for the nontrivial case of a nonvanishing qubit splitting can exceed the wavepacket size. That is, only in the deep strong coupling regime we can expect to observe distinguishable dynamics not only in the qubit occupation, but also in the bosonic field modes. The time evolution of the observed mean displacement $\langle x \rangle$ of the data with nonvanishing qubit spacing depicted in Fig. 3a qualitatively agrees with simulations of the quantum Rabi model in the deep strong coupling regime depicted as lines for large atomic displacements. A more detailed analysis of the real space data in our system again is limited by the finite instrumental resolution of the imaging system.

We have in more detail analyzed the momentum space data obtained by the far-field time of flight imaging, from which both the quasimomentum $q$ and the band index $n_b$ of Bloch bands, with $n_b = 0$ and 1 for momenta measured in the absence of a trapping potential of $p = -2\hbar k + q$ and $p = +2\hbar k + q$ respectively, can be derived (see Methods). Figures 3b and 3c show both the variation of the mean atomic quasimomentum $\langle q \rangle$ and the mean Bloch band occupation $\langle \sigma_x \rangle$, which in the basis of the band eigenstates can be written as $\hat{\sigma}_x = |n_b = 0\rangle\langle n_b = 0| - |n_b = 1\rangle\langle n_b = 1|$, with time along with theory. Remarkably, at small lattice depth, corresponding to a low value of $\omega_q$, we observe a temporally nearly constant value of the Bloch band occupation $\langle \sigma_x \rangle$. This is understood as signal preparation, detection and the system Hamiltonian – the latter in the unusual regime of $g \gg \omega$ being dominated by the interaction term – all are diagonal in the same basis, the eigenbasis of the Pauli matrix $\sigma_x$. In contrast, for larger lattice depth, i.e. with increased $\omega_q$, an oscillatory behavior is observed, as attributed to atomic wavepackets localized in the trap center performing Rabi oscillations between the momentum eigenstates $|\pm 2\hbar k\rangle$ respectively. This is most clearly visible for the data shown by the red squares for $\frac{\omega_q}{2\pi} \cong 3600(40)$ Hz, for which with $\omega_q > \omega$ the dispersive deep strong coupling regime is reached. In general, we observe that the average value of the Bloch band occupation $\langle \sigma_x \rangle$ reduces for large lattice depth, as has been predicted in earlier work [25].

One also finds that near the largest investigated interaction times the experimental data (data points) visible in Figs. 3b and 3c starts to deviate from the theory curves (lines), which were derived based on the quantum Rabi model, as understood from that the edge of the Brillouin zone at $t = \frac{\pi}{2\omega}$ is reached, upon which it becomes relevant that our system realizes a periodic variant of the quantum Rabi model. This is most clearly seen for the data sets recorded with the smallest qubit spacings. Theory predictions based on the periodic quantum Rabi model, which qualitatively reproduce the experimental data also near the band edge, are shown by semi-translucent lines.

**Preparing atoms in qubit eigenstates.** In further measurements, we have prepared atoms in the qubit ground state $|g\rangle$ and excited state $|e\rangle$ respectively formed by the Bloch bands and studied the temporal variation of the qubit population. As described above, the qubit states correspond to coherent superpositions of the momentum picture states $|\pm 2\hbar k\rangle$ respectively, and to prepare these states Bragg transitions were driven with counterpropagating momentum transfer using Raman beams with the corresponding phase difference imprinted. We again start at a vanishing bosonic mode quantum number ($\langle N \rangle = 0$). The initial states $|g, 0\rangle$, $|e, 0\rangle$ prepared in this way correspondingly have different parity[12]. For detection, given that the qubit occupation is encoded in the relative phase of two wavepackets, at the end of the measurement atoms were first adiabatically moved away from the bandgap by chirping the four-photon lattice potential, which remaps the upper and lower bands $|e\rangle$ and $|g\rangle$ to the bare states $|2\hbar k\rangle$ and $|-2\hbar k\rangle$ respectively, and then observing the band population, which allows to determine the population in the corresponding qubit states. Experimental results for the variation of the measured qubit population difference $\langle \sigma_z \rangle = \langle (|e\rangle\langle e| - |g\rangle\langle g|) \rangle$ with time are shown in Fig. 4a. Here the blue dots and yellow triangles

correspond to an initial population in the ground state for a qubit frequency $\omega_q$ of $\omega_q \to 0$ and 1050(10) Hz respectively, and the red squares and green triangles to preparation in the excited state for the corresponding qubit frequencies. In all cases, rapid decay of the population difference $\langle \sigma_z \rangle$ with time is observed, in agreement with theoretical predictions, as understood from the strong coupling of the qubit states with the bosonic field mode leading to a highly entangled nature of the systems eigenstates [12]. We attribute the visible deviation from theory for the largest investigated interaction times of near 700 µs, at which the end of the Brillouin zone is reached, to non-adiabatic transitions occurring in the used experimental band-mapping readout scheme.

We have also determined the variation of the mean excitation number $\langle N \rangle$ on the interaction time, as shown in Fig. 4b for both atoms initially in the qubit ground state $|g\rangle$ (blue dots) and the excited state $|e\rangle$ (orange triangles) respectively. Here an enhancement of the excitation number for atoms initially in the upper qubit state with respect to that when preparing in the lower state is observed. Given that the qubit states are superposition states this demonstrates a dependence of $\langle N \rangle$ on the phase of the initial state, which gives evidence that also at the largest interaction times investigated in Figs. 4a and 4b quantum coherence is preserved. The difference is smaller than theoretical predictions, and we attribute the reduced contrast to the imperfect resolution of our imaging system, which reduces the distinguishability of the diffraction peaks. Figure 4c shows the variation of the difference of the excitation number $\langle N \rangle$ between when preparing the qubit in the ground and excited states respectively on both time and qubit frequency $\omega_q$. This generalizes the results shown in Fig. 4b to different qubit frequency spacings. While for small spacings the sensitivity of the excitation number on the initial state of the qubit is small, at qubit frequencies above $\frac{\omega_q}{2\pi} \approx 1$ kHz a clear difference is visible. While for the simple case of $\omega_q = 0$ the atomic wavepacket superposition oscillating in the trapping potential can be expressed by the Schrödinger cat states $\frac{1}{\sqrt{2}}\left(\left|\frac{ig}{\omega}\right\rangle \pm \left|-\frac{ig}{\omega}\right\rangle\right)$ respectively, for a nonvanishing qubit frequency $\omega_q$ the quantum states become much more complex entangled states. The experimental findings of Fig. 4c demonstrate the phase-dependent behavior of the quantum Rabi dynamics in the deep strong coupling regime, see also the good agreement of the data with theory (Fig. 4d) for comparison.

**Discussion**

Our experiment demonstrates that quantum Rabi physics at unpreceded high coupling strength can be realized with ultracold atoms in optical lattices using solely the spatial degrees of freedom. In our approach, a two-level system has been encoded in the occupation of Bloch bands, interacting with a bosonic mode implemented by harmonic motion in a dipole trap. The characteristic dynamics at these parameter regimes has been mapped out.

For the future, extensions of this work can include quantum information processing based on qubits encoded in the vibrational dynamics of cold atoms in engineered superpositions of periodic lattices and slowly varying dipole trapping potentials. This is reminiscent of work done in the phase space of superconducting qubit systems, albeit with stronger coupling strengths [24,26]. For coupling of different qubits, digital techniques, alternating between tightly confined interaction and qubit manipulation periods following here demonstrated techniques can be envisioned [27]. It also will be interesting to extend the present work to longer interaction times, as to study predictions of the periodic quantum Rabi model and observe collapse and revival patterns of the initial state [21]. Other interesting future work includes the search for phase transitions of the spin-boson model [28,29].

**Methods**

**Experimental setup and procedure**

Our experimental apparatus, see also the schematics shown in Fig. 1b of the main text, is a modified version of a setup used in earlier works [30]. Inside a vacuum apparatus, cold rubidium atoms ($^{87}$Rb) collected in a magneto-optic trap are loaded into the dipole trapping potential induced by a beam focused to 42 µm diameter derived from a $CO_2$-laser operating near 10.6 µm wavelength. The atoms are evaporatively cooled to quantum degeneracy by lowering the depth of the dipole potential. In the final stages of the cooling, a magnetic quadrupole field is activated, which allows to generate a spin-polarized Bose-Einstein condensate in the $m_F = 1$ component of the $F = 1$ hyperfine ground state. To keep interaction effects small, here we work with small condensate numbers of typically 2500 atoms, as achieved by ramping the depth of the trapping potential in evaporative cooling to lower values than needed to achieve condensation to reduce the number of confined atoms. Subsequently, the dipole trapping potential is ramped up adiabatically within 250 ms to reach the desired values of the trapping frequency $\omega$, see also the main text, for simulation of the quantum Rabi model. Typical beam powers are 32 W, 30 mW, and 100 mW during loading, the final stage of evaporative cooling, and quantum Rabi manipulation phases, respectively. In the latter phase, atoms remain well confined in the center of the Gaussian beam, in a range where the dipole trapping potential can well be described as a harmonic potential. The anharmonicity, defined as the difference of the dipole potential expected to be imprinted by a Gaussian laser beam and a harmonic potential, for the experimentally relevant parameter regime is below 0.25%. The estimated decoherence rate from photon scattering from the mid-infrared trapping laser beam in the quantum Rabi manipulation phase is 1.7x10$^{-5}$/s, i.e., is negligible. In practice, decoherence will be determined by scattering from the Raman beams and atomic interaction effects.

The method used to generate a high spatial harmonic lattice potential of periodicity $\frac{\lambda}{4}$, where $\lambda \simeq$ 783.5 nm (which is detuned 3.5 nm from the rubidium D2-line) denotes the wavelength of the driving laser beams, relies on four-photon Raman processes [22]. The transitions are driven in a three-level configuration with two stable ground states $|1\rangle$ and $|2\rangle$ and one spontaneously decaying excited state $|3\rangle$ by a beam of frequency $\omega_{lat}$ and two counterpropagating superimposed beams of frequencies $\omega_{\text{lat}} + \Delta\omega_{\text{lat}}$ and $\omega_{\text{lat}} - \Delta\omega_{\text{lat}}$. The $m_F = -1$ and $m_F = 0$ components of $5S_{1/2}, F = 1$ constitute the used ground states and the $5P_{3/2}$ manifold serves as the excited state of the three-level configuration. Atomic momentum is exchanged with the driving light field in units of four-photon momenta, which is a factor two above that of the relevant processes in a usual standing wave lattice induced by two-photon processes. Correspondingly, the spatial periodicity of the induced potential is a factor two smaller and equals $\frac{\lambda}{4}$.[22] In the experiment, we typically use a magnetic bias field of 1 G to remove the degeneracy of Zeeman sublevels and a frequency difference $\frac{\Delta\omega_Z}{2\pi} \simeq 1.6$ MHz.

In the experimental sequence, following the ramping up of the $CO_2$-laser beam intensity to the desired harmonic trapping frequency, atoms are prepared near the first avoided crossing of the lattice band structure (Fig. 1d) by means of Bragg diffraction. For the used lattice with spatial periodicity $\frac{\lambda}{4}$ usual Bragg diffraction, transferring momentum in units of two-photon momenta, can be used to prepare atoms at the position of the first band crossing. For the experimental data shown in Fig. 4, with qubit states $|g\rangle$ and $|e\rangle$ respectively as the initial states, two simultaneously performed Bragg pulses with opposite directions of the momentum transfer were used with the relative phase of the pulses allowing to set the desired qubit initial state. Following preparation, atoms were left in the desired combined potential of lattice and harmonic trapping for quantum Rabi manipulation for a variable interaction time.

Subsequent detection of the atomic cloud was performed after extinguishing both the lattice and the dipole trapping beams. For this, absorption imaging of the atomic sample was employed onto a sCMOS camera. During the experiments described in the main text, both measurements probing the real-space distribution are carried out by probing directly following the experiment, as well as far-field time-of-flight imaging probing the momentum distribution were performed. For an analysis of measurements of the rms displacement $\langle x^2 \rangle$ of the atomic cloud from the trap center, the experimental image data was first deconvoluted by the point spread function of the imaging system (of near 6.5 µm instrumental resolution) determined in an independent measurement before analysis of this moment from a series of measurement. Example images after deconvolution are shown in Fig. 5a.

The momentum $p$ measured in the absence of a trapping potential maps onto the quasimomentum $q$ and the band index $n_b \in \{0,1\} \to \{-2\hbar k, 2\hbar k\}$ mapping the basis states of our qubit state via $p = q + 2\hbar k(2n_b - 1)$. Example time of flight images to obtain the momentum $p$ and subsequently quasimomentum $q$ and band index $n_b$ are shown in Fig. 5b.

**Theoretical methods**

The single-particle Hamiltonian for a cloud of ultracold atoms is described by the sum of a harmonic part, which includes the kinetic energy of the atoms and the harmonic trap, and a periodic potential,

$$\hat{H} = \frac{\hat{p}^2}{2m} + \frac{m\omega^2}{2}\hat{x}^2 + \frac{V}{2}\cos(4k\hat{x}), \qquad (2)$$

where $\hat{p} = -i\hbar \frac{d}{dx}$ and $\hat{x}$ are momentum and position of an atom of mass $m$, respectively. If we write this Hamiltonian in the Bloch basis function $\langle x|\phi_n(q)\rangle = e^{iqx/\hbar}e^{-i2kx}e^{i4nkx}$ and we project to the two lowest energy bands, it is recast into

$$\hat{H} = \frac{\hat{q}^2}{2m} + \frac{m\omega^2}{2}\hat{x}^2 + \frac{2\hbar k}{m}\begin{pmatrix}1 & 0\\ 0 & -1\end{pmatrix}\hat{q} + \frac{V}{4}\begin{pmatrix}0 & 1\\ 1 & 0\end{pmatrix}. \qquad (3)$$

Defining creation and annihilation operators $\hat{a} = \sqrt{\frac{m\omega}{2\hbar}}(\hat{x} + \frac{i}{m\omega}\hat{q})$ and $\hat{a}^\dagger = \sqrt{\frac{m\omega}{2\hbar}}(\hat{x} - \frac{i}{m\omega}\hat{q})$, while rotating the qubit (band index) with the unitary operator $U = \frac{1}{\sqrt{2}}\begin{pmatrix}1 & 1\\ 1 & -1\end{pmatrix}$, and defining the Pauli matrices in the rotated basis as

$$\sigma_x = |n=0\rangle\langle n=0| - |n=1\rangle\langle n=1|$$

$$\sigma_z = |n=1\rangle\langle n=0| + |n=0\rangle\langle n=1|$$

the total system Hamiltonian is the one given in Eq. (1).

In our experimental sequence, the prepared initial states correspond to quite highly energetic states of the system Hamiltonian, as can be seen in Fig. 6, which shows the numerically determined occupation probability of the states for atoms prepared in a $|-2\hbar k\rangle$ momentum state and the bosonic vacuum state, corresponding to typical initial conditions. The initial state $|-2\hbar k\rangle$ corresponds to a superposition of two qubit states, such that here both eigenstates with negative and positive parity (see Fig. 6a and Fig. 6b respectively) are populated.

In what follows, we would like to understand the occupation number, or number of photons, in an experiment starting with the ground state of a cavity mode with frequency $\omega$ in a Rabi model, where the frequency of the qubit is set to zero, i.e. $\omega_q = 0$, and the coupling strength of the cavity and qubit is given by $g$,

$$\begin{aligned} H_{\omega_q=0} &= \omega\left(N + \frac{1}{2}\right) + g\sigma_x(a + a^\dagger) \\ &= \omega\left(a^\dagger + \frac{g\sigma_x}{\omega}\right)\left(a + \frac{g\sigma_x}{\omega}\right) + \frac{\omega}{2} - \frac{g^2}{\omega} \end{aligned} \quad (4)$$

As it can be directly seen from the second line, this Hamiltonian is diagonal with the displaced cavity operators

$$b = a + \frac{g\sigma_x}{\omega} = a + \alpha,$$

where the displacement operator is given by

$$D(\alpha) = e^{\alpha a^\dagger - \alpha^* a} = e^{-\frac{|\alpha|^2}{2}} e^{\alpha a^\dagger} e^{-\alpha^* a} \quad ,$$

for a general $\alpha$ parameter.

The action of the time evolution of a displaced Hamiltonian for a cavity mode on the vacuum, i.e., no excitations, is given by

$$e^{iH(\alpha)t}|0\rangle = D(-\alpha)e^{iH(0)t}D(\alpha)|0\rangle = e^{\frac{i\omega t}{2}}e^{\frac{ig^2 t}{\omega}}D(-\alpha)|e^{-i\omega t}\alpha\rangle$$

$$= e^{\frac{i\omega t}{2}}e^{\frac{ig^2 t}{\omega}}e^{\text{Im}(|\alpha|^2 e^{i\omega t})}D[\alpha(e^{-i\omega t}-1)]|0\rangle.$$

From this expression, we can derive the expectation value in the number of excitations for an interaction time $t$

$$\langle N \rangle = |\alpha(t)|^2 = 4|\alpha|^2 \sin^2\left(\frac{\omega t}{2}\right). \tag{5}$$

Correspondingly, with $\alpha = \frac{g}{\omega}$, the maximum number of the expectation value $\langle N \rangle$ is given by

$$N_{\max} = 4|\alpha|^2 = \frac{4g^2}{\omega^2}. \tag{6}$$

**Comparison to fluxonium qubit system**

Finally, we remark that the here relevant Hamiltonian Eq. 2 maps onto Hamiltonians reached with superconducting fluxonium systems. Specifically, see Eq. 1 of the quasicharge qubit system of Ref. 24, which reads:

$$H = E_C\left(\frac{Q}{2e}\right)^2 + \frac{1}{2}E_L\varphi^2 - E_J\cos(\varphi - \varphi_{\text{ext}}), \tag{7}$$

with Q as the charge and $\varphi$ as the superconducting phase difference. Further, $E_C$ denotes the charging energy, $E_L$ the inductive energy, $E_J$ the Josephson energy, and $\varphi_{\text{ext}}$ an external phase. By separating the time-dependent Schrödinger equation obtained with the Hamiltonian of Eq. 7 into slow and fast varying parts respectively and substitution of $\varphi = 4kx$ we find that resulting equation of motion can be written in terms of an effective Hamiltonian that up to a basis transformation directly maps onto Eq. 3 for

$$E_C/\hbar = \frac{2\hbar k^2}{m}, \quad E_J/\hbar = \omega_q, \quad E_L/\hbar = \frac{m\omega^2}{16\hbar k^2}, \quad \hbar g = (8E_L E_C^3)^{\frac{1}{4}}, \tag{8}$$

For the specific case of an external phase $\varphi_{\text{ext}} = \pi$ one also finds that with these identifications Eq. 2 and equation 1 of Ref. 24 are akin. This allows us to directly compare the energy scales given in Ref. 24 to the parameters used here:

$$\frac{g}{\omega} \simeq 1.91, \qquad \frac{\omega_q}{\omega} \simeq 2.42.$$

The here derived value of the (normalized) coupling strength $g/\omega$ of this superconducting system is above that of earlier works explicitly studying quantum Rabi physics in superconducting systems, and below the corresponding values of both the ion trapping work of Ref. 20 and of the present work.

**Acknowledgements**

We thank one of the referees for pointing out the connection of the Hamiltonian of the used cold atom system to the fluxonium system of Ref. 24. Support by the DFG within the project We 1748-24 (642478) (M.W.), the focused research center SFB/TR 185 (277625399) (M.W.) and the Cluster of Excellence ML4Q (EXC 2004/1 – 390534769) (M.W.), the QMiCS (820505) (E.S.) and OpenSuperQ (820363) (E.S.) projects of the EU Flagship on Quantum Technologies, the National Natural Science Foundation of China (NSFC) (12075145) (E.S.), the Shanghai Government grant STCSM (2019SHZDZX01-ZX04) (E.S.), the Spanish Government PGC2018-095113-B-I00 (MCIU/AEI/FEDER, UE) (E.R.), the Basque Government IT986-16 (E.R.), and the FET Open Quromorphic (E.S.) and EPIQUS (E.S.) EU projects is acknowledged.

**Author Contributions**

J.K. and G.H. conducted the experiment and analyzed the data. T.O. conducted early experimental groundwork. E.R. did numerical and analytical studies and simulations. E.S. and M.W. planned the project. All authors contributed to the writing of the manuscript and the data interpretation.

**Competing interests**

The authors declare no competing interests.

**Figures**

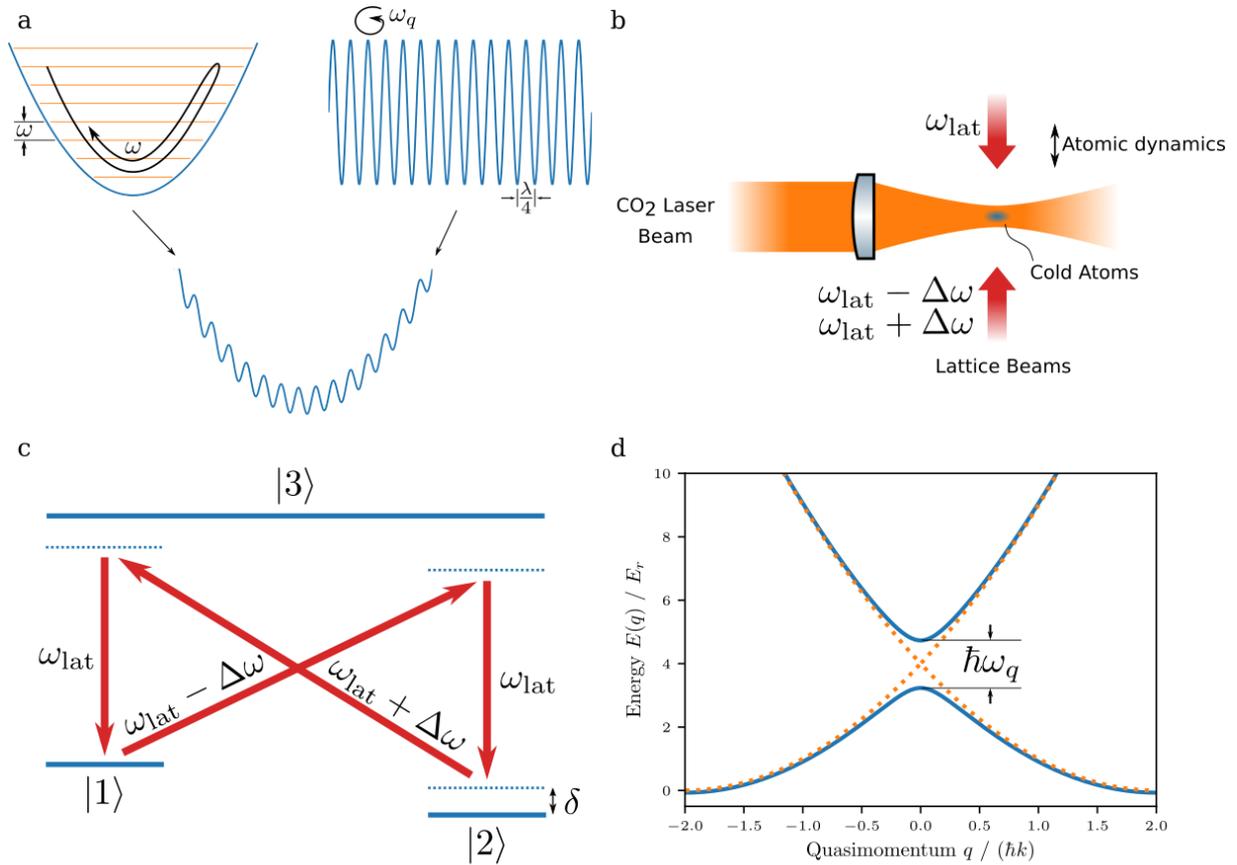

**Fig. 1: Experimental schematic**. **a**, Atoms are exposed to the combined potential obtained by superimposing a harmonic trapping potential (left) generated by a focused $CO_2$-laser beam and a lattice potential of spatial periodicity $\frac{\lambda}{4}$ (right). The relevant oscillatory modes, of frequency $\omega$ for the oscillation in the harmonic trapping potential and $\omega_q$ for oscillation at the first band gap of the lattice, are indicated. For atoms moving in the combined potential, the two modes are very strongly coupled. **b**, Schematic of the experimental setup along with the optical frequency components in the optical lattice beams used to synthesize the four-photon lattice potential of periodicity $\frac{\lambda}{4}$, see **c** for the coupling scheme. **d**, Dispersion relation of rubidium atoms in the lattice (blue) versus the atomic quasimomentum along with the dispersion of free atoms in states $|2\hbar k + q\rangle$ and $|-2\hbar k + q\rangle$ (orange dotted). At the position of the crossing, atoms in the lower and upper band correspond to states $|g\rangle$ and $|e\rangle$ respectively of the two-level qubit system.

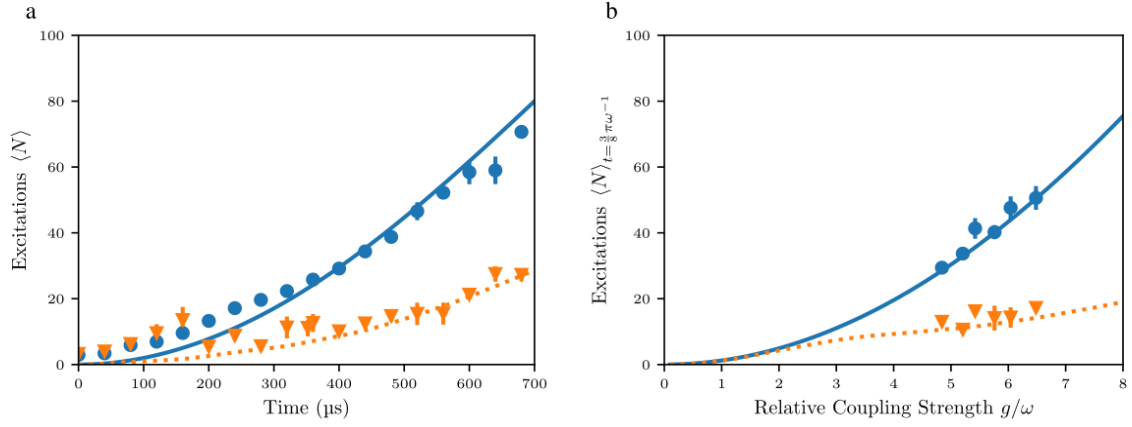

**Fig. 2: Creating system excitations**. **a**, Variation of the number of excitation quanta $\langle N \rangle$ in the potential versus interaction time for a coupling $\frac{g}{2\pi} = 2275(23)$ Hz and a bosonic mode frequency $\frac{\omega}{2\pi} = 346(7)$ Hz, corresponding to a relative coupling strength $\frac{g}{\omega} = 6.58(7)$, i.e., far in the deep strong coupling regime. The used qubit spacings were $\frac{\omega_q}{2\pi} = 586(6)$ Hz (blue dots) and 5200(50) Hz (orange triangles). The lines are theory. Atoms for this measurement are prepared in the momentum state $|-2\hbar k\rangle$ in the center of the harmonic trapping potential. **b**, Variation of the number of excitations on the relative coupling strength $\frac{g}{\omega}$. Here a fixed interaction time of $\frac{3}{8}\frac{\pi}{\omega}$ was used, and blue dots and orange triangles correspond to qubit spacings $\frac{\omega_q}{2\pi} = 590(6)$ Hz and 5850(60) Hz respectively. The visible error bars denote the statistical uncertainties.

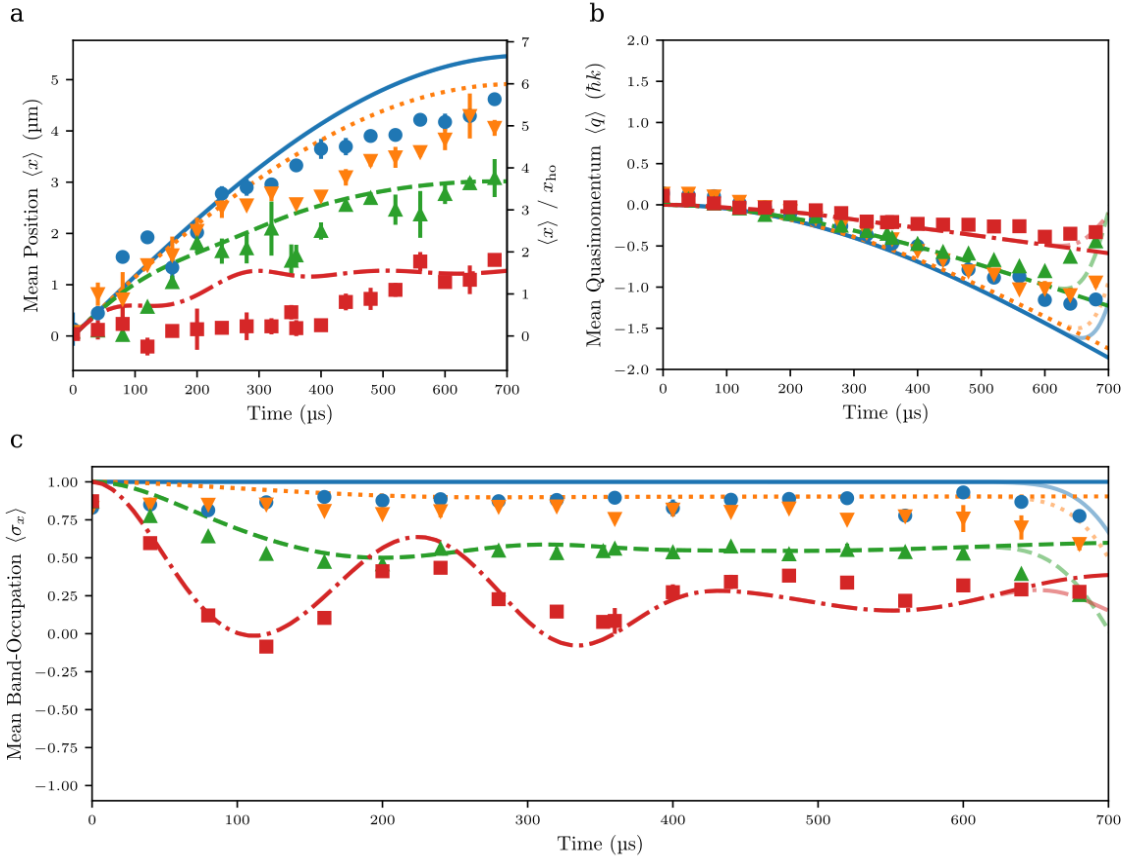

**Fig. 3: Revealing the time evolution of system parameters. a**, Variation of the mean value of the atomic cloud position $\langle x \rangle$ on the interaction time. On the right hand scale, this quantity is given in units of the harmonic oscillator length $x_{\text{ho}}$. Experimental parameters were a coupling $\frac{g}{2\pi} = 2275(23)$ Hz and a bosonic mode frequency $\frac{\omega}{2\pi} = 346(7)$ Hz. Further, the used qubit spacing $\frac{\omega_q}{2\pi}$ was 0 Hz (blue dots), 586(6) Hz (orange triangles), 1660(20) Hz (green upside-down triangles), and 3600(40) Hz (red squares) respectively. The system is initially prepared in the momentum state $|-2\hbar k\rangle$. Theory results for the quantum Rabi and the periodic quantum Rabi models are represented by non-transparent and semi-transparent lines, respectively. **b**, Time evolution of the observed mean atomic quasimomentum $\langle q \rangle$ and **c**, the Bloch band occupation $\langle \sigma_x \rangle$ for corresponding values of the qubit spacing. The visible error bars denote the statistical uncertainties.

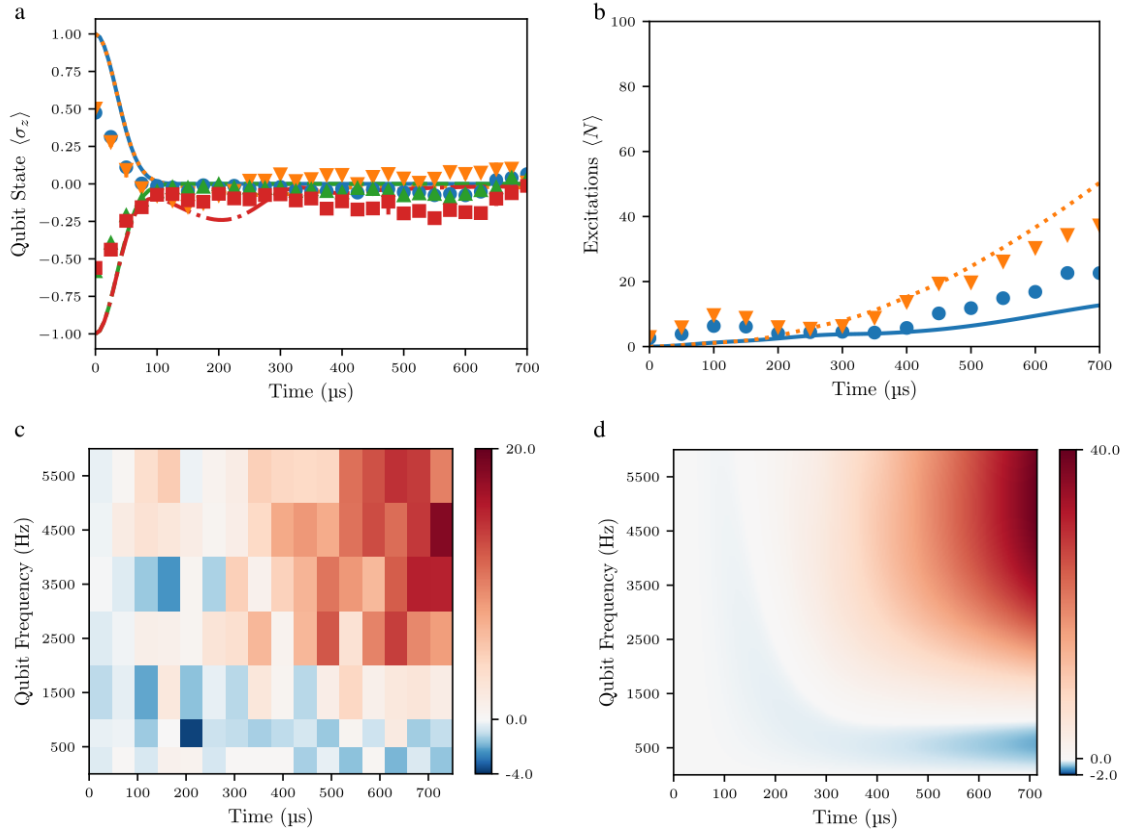

**Fig. 4: Preparing atoms in qubit states. a**, Time evolution of the qubit excitation $\langle\sigma_z\rangle$ following preparation of atoms in lower (orange triangles) and upper (red squares) states of the two-state system for a qubit spacing $\frac{\omega_q}{2\pi} = 1050(10)$ Hz. For comparison, the blue dots and green upside-down triangles correspond to measurements with $\omega_q = 0$. The lines are theory. $\frac{g}{\omega} = 6.50(5)$ in all measurements. **b**, Variation of the mean excitation number <N> on the interaction time for atoms initially prepared in the ground (blue dots) and the excited qubit states (orange triangles) respectively for $\frac{\omega_q}{2\pi} = 4660(50)$ Hz, along with theory (lines). **c**, Experimental data for the difference in the mean excitation number $\langle N\rangle_{|e\rangle} - \langle N\rangle_{|g\rangle}$ observed when preparing atoms initially in qubit excited and ground states respectively on both interaction time and qubit spacing represented in color code. **d**, Corresponding theory expectations.

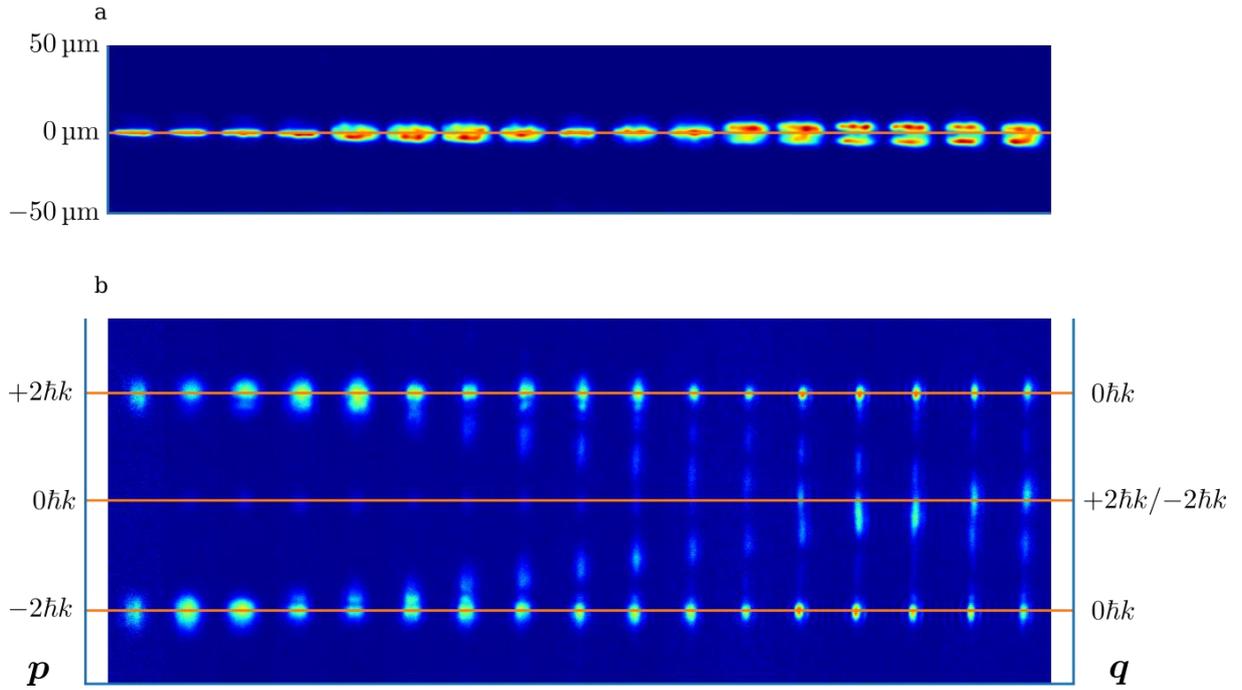

**Fig. 5: Examples of obtained image data. a**, Series of real space images for the system initially prepared in the qubit excited state of the quantum Rabi Hamiltonian, with $\frac{\omega_q}{2\pi} = 2380(40)$ Hz and $\frac{g}{\omega} = 6.5(5)$, after deconvolution of the obtained raw absorption imaging data of the atomic ensemble accounting for the point spread function of the used optics. From left to right the interaction time increases in steps of 50 µs. Despite the limited instrumental optical resolution, a splitting up of the atomic cloud is observed. **b,** Series of time-of-flight images of the atomic cloud, for the same system state as in subfigure **a**. The numbers on the left-hand side represent the measured atomic momentum $p$, while the on the right-hand side the corresponding quasimomentum $q$ is given.

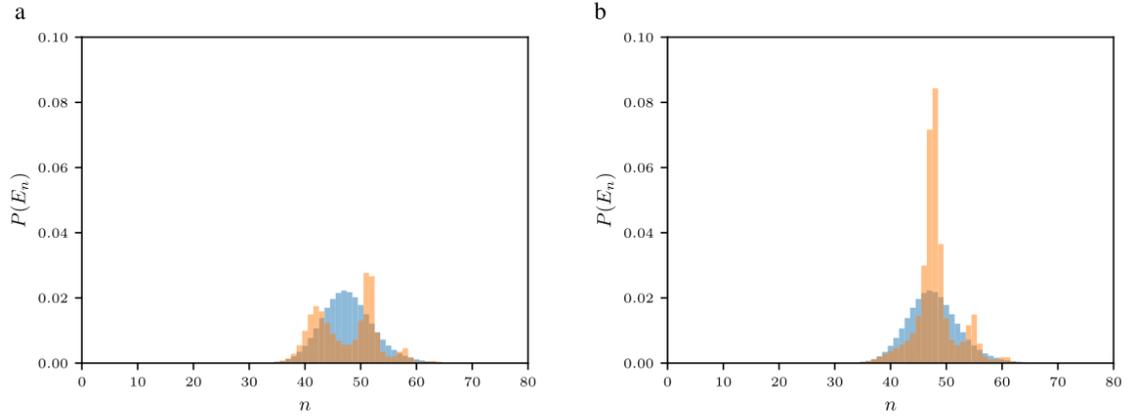

**Fig. 6: Simulation of the initial system state. a**, Population distribution of negative parity eigenstates of the system when initially prepared in the $|-2\hbar k\rangle$ momentum state, and the vacuum field mode (<N>=0), versus the excitation number *n*. The case of $\frac{\omega_q}{2\pi} = 0$ Hz is shown in blue color, and the case of $\frac{\omega_q}{2\pi} = 1000$ Hz is shown in orange. **b,** Corresponding population distribution of the positive parity eigenstates.